\documentclass[a4paper,11pt]{article}

\usepackage{latexsym}
\usepackage{amsfonts} 
\usepackage[mathscr]{euscript}
\usepackage{amsmath,mathrsfs,bm,amssymb,color}
\usepackage{fancybox}

\title{\textsf{ Monotonicity of the polaron energy}}
\date{\empty}
\author{
Tadahiro Miyao\\ 
 {\it Department of Mathematics,}
{\it Hokkaido University,}\\
{\it Sapporo 060-0810, Japan}\\
e-mail:
 miyao@math.sci.hokudai.ac.jp
}

\newcommand{\one}{{\mathchoice {\rm 1\mskip-4mu l} {\rm 1\mskip-4mu l}
{\rm 1\mskip-4.5mu l} {\rm 1\mskip-5mu l}}}
\newcommand{\h}{\mathfrak{h}}

\newcommand{\ex}{\mathrm{e}}
\newcommand{\D}{\mathrm{dom}}

\newcommand{\im}{\mathrm{i}}
\newcommand{\Fock}{\mathfrak{F}}
\newcommand{\Ffin}{\mathfrak{F}_{\mathrm{fin}}}

\newcommand{\dG}{\mathrm{d}\Gamma}

\newcommand{\ran}{\mathrm{ran}}

\newcommand{\la}{\langle}
\newcommand{\ra}{\rangle}

\newcommand{\BbbR}{\mathbb{R}}
\newcommand{\BbbN}{\mathbb{N}}

\newcommand{\BbbC}{\mathbb{C}}
\newcommand{\vepsilon}{\varepsilon}
\newcommand{\vphi}{\varphi}

\newcommand{\Pf}{P_{\mathrm{f}}}
\newcommand{\Nf}{N_{\mathrm{f}}}

\newcommand{\Cone}{\mathfrak{p}}

\newcommand{\dm}{\mathrm{d}}

\newcommand{\no}{\nonumber \\}

\def\Sumoplus{\sideset{}{^{\oplus}_{n\ge 0}}\sum}

\begin{document}

\newtheorem{define}{Definition}[section]
\newtheorem{Thm}[define]{Theorem}
\newtheorem{Prop}[define]{Proposition}
\newtheorem{lemm}[define]{Lemma}
\newtheorem{rem}[define]{Remark}
\newtheorem{assum}{Condition}
\newtheorem{example}{Example}
\newtheorem{coro}[define]{Corollary}

\maketitle

\begin{abstract}
In condensed matter physics, the polaron has been fascinating subject.
It is described by the Hamiltonian of H. Fr\"ohlich.
In this paper, the Fr\"ohlich Hamiltonian is investigated from a
 viewpoint of operator inequalities  proposed  in
 \cite{Miyao2}. This point of view clarifies
the monotonicity of polaron energy, i.e., denoting the lowest energy of 
the Fr\"ohlich Hamiltonian  with the ultraviolet cutoff $\Lambda$ by
 $E_{\Lambda}$, we prove $E_{\Lambda}>E_{\Lambda'}$ for $\Lambda<\Lambda'$.

\end{abstract} 

\section{Introduction}

Let us consider an  electron in an ionic crystal.
Through the Coulomb interaction, the electron polarizes the lattice
all  around itself. The  moving electron carries the lattice
polarization with it.  Hence it is natural to regard the moving electron and its
accompanying
 distortion  field as one object, called the  polaron. 
This system is described by the  Hamiltonian of H. Fr\"ohlich \cite{HFroehlich}
and of interest in condensed matter physics.
As a model for a nonrelativistic particle coupled with the bosonic
field,  this Hamiltonian has been widely  studied by many authors.
Although quite a number of 
literatures
 can be found, we refer to  \cite{Dev, Feynman2} as accessible works.
In general, there are two approaches to investigate  the Fr\"ohlich
Hamiltonian.
One is based on the Feynman's path integral approach \cite{DoVa,
Feynman, Spohn1},
and other one is standard operator theoretical methods \cite{
GeLowen, LGross2,LLP,  Sloan1, Sloan2}.
Nowadays we already reaped a rich harvest from the Fr\"ohlich 
Hamiltonian by both approaches. Nevertheless  it is still an attractive
subject \cite{BG,DM,  FLST, FLS, FLS2, GHW, GM, LT1, MS,
Moller2,  Spohn2}.

The Fr\"ohlich Hamiltonian is formally given  by 
\begin{align}
H=-\frac{1}{2}\Delta_x -\sqrt{\alpha} \lambda_0 \int_{\BbbR^3}\dm k \frac{1}{|k|}[\ex^{\im k\cdot
 x} a(k)+\ex^{-\im k\cdot x }a(k)^*]+\int_{\BbbR^3}\dm k a(k)^* a(k).\label{FormalDef}
\end{align} 
Since the coupling  function $1/|k|$ is not square integrable, the interaction term in (\ref{FormalDef})
is not well-defined methematically. Here recall that
$a(f)=\int_{\BbbR^3}\dm k  f(k)^* a(k)$
is well-defined only if $f$ is square integrable.
In order to define $H$ rigorously, we usually employ  the following procedure.
We first introduce  the Hamiltonian with an ultraviolet(UV) cutoff by (\ref{FullHami}).
Then we can show that there exists a self-adjoint operator $H$ such that 
 $H_{\Lambda}$ converges to  $H$ in the strong resolvent sense as
 $\Lambda\to \infty$ \cite{ Eck, JFroehlich1, JFroehlich2, LGross2,
 Nelson, Sloan1, Sloan2}.
In this way, we can define the Fr\"ohlich Hamiltonian as a limiting operator, and 
clarify   a mathematical meaning of the formal definition  (\ref{FormalDef}).
Our next problem is  the spectral analysis of the Fr\"ohlich Hamiltonian
$H$.
 In \cite{JFroehlich1, JFroehlich2}, J. Fr\"ohlich studied  the spectral properties 
of the Hamiltonian at a fixed total momentum after removal of 
UV cutoff. See also \cite{GeLowen, HHS, Moller2, Pizzo}. 
In these papers,  the existence of a ground state was also proven.
Moreover the uniqueness of the ground state was studied by applying the 
Perron- Frobenius theorem.

The Fr\"ohlich Hamiltonian  is a reasonable example of  an application of the
Perron-Frobenius theorem. There are several beautiful works on this
theorem. Many of these have been developed in order to show the
uniqueness of the ground state in the nonrelativistic quantum field theory
\cite{Faris, LGross, LGross1, Sloan1}.
Of course, the Fr\"ohlich Hamiltonian  has been a target for this theorem as well, 
but known as a tough problem because of   difficulties  comming from the
removal of UV cutoff.
In the previous work \cite{Miyao}, the author proved the uniqueness of the ground
state for the Fr\"ohlich Hamiltonian.  Some new operator inequalities
which will be  discussed in later sections
 played some  important roles.
Especially  an operator monotonicity was essential for the
proof. 
Our main purpose in this paper is to search for further applications of the operator inequalities to the
Fr\"ohlich Hamiltonian. 
We will  prove the monotonicities of the polaron energy. Throughout our
proofs, we will see how useful our operator inequalities are.

The paper is organized as follows: In Section \ref{MainResults}, we
define some   models and display our results. In Section
\ref{SecMono}, we explain our strategy of proof as an abstract theorem.
 Section \ref{2ndQuant} deals with the second quantization and Sections
 \ref{ProofEMW} and \ref{ProofLocalPr} with the proofs of main results.
In Appendix \ref{SDCA}, we review some basic facts about the operator inequalities. In Appendix \ref{LocalEnergyInq}, we show a useful energy inequality.
\medskip\\
 {\bf Acknowledgements}\\
I would like to thank Herbert Spohn for useful discussions.
This work was  supported by KAKENHI (20554421).

\section{Main results}\label{MainResults}

\setcounter{equation}{0}

In this section, we define several Hamiltonians  and display our main results.
Mathematical definitions of second quantized operators 
will be given in \S \ref{2ndQuant}. If readers are unfamiliar with the
second quantization, we recommend them to read \S \ref{2ndQuant} first.

\subsection{Definition of Hamiltonians with  the  ultraviolet cutoff}\label{DefHamiUV}

Let $\Fock $ be the Fock space over $L^2(\BbbR^3)$.
For each $\Lambda>0$, we define a linear operator
$H_{\Lambda}$
living in $L^2(\BbbR^3_x)\otimes \Fock$ by
\begin{align}
H_{\Lambda}=-\frac{1}{2}\Delta_x-\sqrt{\alpha}\lambda_0
\int_{|k|\le \Lambda}\dm k\, \frac{1}{|k|}\big[\ex^{\im k\cdot x}a(k)+
\ex^{-\im k\cdot x}a(k)^*
\big]
+\Nf,\label{FullHami}
\end{align}
where $\Delta_x$ is the  Laplacian on $L^2(\BbbR^3_x)$, 
$\sqrt{\alpha}$ is the electron-phonon coupling strength
 and
$\lambda_0=2^{1/4}(2\pi)^{-1}$. 
$a(k)$ and $a(k)^*$ are the phonon annihilation- and creation operators
with commutation relations
\begin{align}
[a(k), a(k')^*]=\delta(k-k'),\ \ [a(k), a(k')]=0.
\end{align} 
$\Nf$ is the number operator formally expressed as
$\Nf=\int_{\BbbR^3}\dm k a(k)^* a(k)$.
(The complete definitions of these operators will be given  in \S
\ref{2ndQuant}.)
Let us denote the smeared annihilation- and creation operators by 
$a(f)$ and $a(f)^*$ for $f\in L^2(\BbbR^3)$.
These are  expressed as 
\begin{align*}
a(f)=\int_{\BbbR^3}\dm k\, \overline{f(k)} a(k),\ \ \ a(f)^*=\int_{\BbbR^3}\dm k\,
 f(k) a(k)^*
\end{align*}
respectively. 
Then, by the standard  bound
\begin{align}
\|a(f)^{\#}(\Nf+\one)^{-1/2}\|\le \| f\|\label{StandardBound}
\end{align} 
and the Kato-Rellich theorem \cite{ReSi2}, 
$H_{\Lambda}$ is self-adjoint on $\D(\Delta_x)\cap \D(\Nf)$ and bounded
from below.
$H_{\Lambda}$ is  called the Hamiltonian with an UV cutoff
$\Lambda$.

Let $P_{\mathrm{tot}}$ be the total momentum operator defined by
\begin{align}
P_{\mathrm{tot}}=-\im \nabla_x+\Pf.
\end{align} 
Here $\Pf$ is the phonon momentum operator given by
$\Pf=\int_{\BbbR^3}\dm k ka(k)^* a(k)$.
Let $\mathcal{F}_x$ be the Fourier
 transformation
on $L^2(\BbbR^3_x)$ and let $U=\mathcal{F}_x \ex^{\im x\cdot \Pf}$. 
 Then we can   see 
\begin{align}
U P_{\mathrm{tot}} U^*=\int^{\oplus}_{\BbbR^3} P\, \dm P.
\end{align} 
Moreover  one has
\begin{align}
UH_{\Lambda}U^*=\int^{\oplus}_{\BbbR^3} H_{\Lambda}(P)\, \dm P. \label{Fiber}
\end{align} 
Each $H_{\Lambda}(P)$ is given by 
\begin{align}
H_{\Lambda}(P)=\frac{1}{2}(P-\Pf)^2-\sqrt{\alpha}\lambda_0
\int_{|k|\le \Lambda}\dm k\, \frac{1}{|k|}[a(k)+a(k)^*]+\Nf. \label{FixHamiCutt}
\end{align} 
Then, by (\ref{StandardBound}) and the  Kato-Rellich theorem,
$H_{\Lambda}(P)$ is self-adjoint on $\D(\Pf^2)\cap \D(\Nf)$, bounded
from below.
The self-adjoint operator (\ref{FixHamiCutt}) is called the Hamiltonian
 at  a fixed total momentum $P$.

\subsection{ Removal of ultraviolet cutoff}
One of basic   problems is removal of UV cutoff $\Lambda$. Namely we
would like to take a limit  $\Lambda\to \infty$ in (\ref{FullHami}) and
(\ref{FixHamiCutt}). However, since $1/|k|$ is not square
integrable,
the interaction terms are not well defined in this limit.
Therefore the standard perturbation methods, like the  Kato-Rellich theorem,
can not be applicable to define Hamiltonians without  UV
cutoff.
In other words,  we face  a singular perturbation  problem.
Fortunately this problem was already solved by several authors.
Here we just state the results.

\begin{Prop}\label{LimitHami}
One obtains the following.
\begin{itemize}
\item[{\rm (i)}] There exists a unique semibounded self-adjoint operator $H$
such  that $H_{\Lambda}$ converges to $H$ in strong resolvent sense as
	   $\Lambda\to \infty$.
\item[{\rm (ii)}] For each $P\in \BbbR^3$,  there exists a unique semibounded self-adjoint operator $H(P)$
such  that $H_{\Lambda}(P)$ converges to $H(P)$ in strong resolvent sense as
	   $\Lambda\to \infty$.
\end{itemize} 
\end{Prop} 
As to the proof of Proposition \ref{LimitHami}, we refer to \cite{GeLowen,Nelson}.

\subsection{Results }
Let
\begin{align} 
E_{\Lambda}=\inf \mathrm{spec}(H_{\Lambda}),\ \  E=\inf
 \mathrm{spec}(H).
\end{align} 
Similarly let
\begin{align} 
E_{\Lambda}(P)=\inf \mathrm{spec}(H_{\Lambda}(P)),\ \  E(P)=\inf
 \mathrm{spec}(H(P)).
\end{align} 
We consider  the following question:
\begin{center}
{\it
How do $E_{\Lambda}$ and $E_{\Lambda}(P)$ behave as  functions of $\Lambda$?
}
\end{center} 
Of course, we already know that 
\begin{align}
\lim_{\Lambda\to \infty}E_{\Lambda}=E,\ \ \lim_{\Lambda\to \infty}E_{\Lambda}(P)=E(P)
\end{align} 
 by Proposition \ref{LimitHami}.
Hence our real motive behind the question  is that  we wish to know more detailed
information about the behaviors of $E_{\Lambda}$ and $E_{\Lambda}(P)$.

In subsequent  sections, we will show the following theorems.

\begin{Thm}\label{EnergyMonoWeak}
For each $P\in \BbbR^3$, $E_{\Lambda}(P)$ is monotonically decreasing in $\Lambda$.
\end{Thm}

If the total momentum $P$ is small enough, one can
obtain  a stronger result.

\begin{Thm}\label{EnergyMonoPolaron}
 For each $P\in \BbbR^3$ with $|P|<\sqrt{2}$, $E_{\Lambda}(P)$ is strictly decreasing in $\Lambda$.
\end{Thm} 

\begin{rem}
{\rm
Theorems \ref{EnergyMonoWeak} and \ref{EnergyMonoPolaron} suggest the
 polaron at a fixed total momentum is more stable, the larger we take
 the UV cutoff. Hence the polaron  is energetically  most
 stable if the UV cutoff is removed.
}
\end{rem}

Combining the fact  $E_{\Lambda}=E_{\Lambda}(0)$ by (\ref{Fiber}) and the well-known property
$E_{\Lambda}(0)\le E_{\Lambda}(P)$, one has the following corollary.

\begin{coro} 
$E_{\Lambda}$ is strictly decreasing in $\Lambda$.
\end{coro}

\begin{rem}
{\rm

In \cite{Moller1, Moller2}, Moller considered a similar model with a smooth UV cutoff function $\hat{\vphi}(k)$ which is strictly positive.
He showed some  monotonicities of the ground state energy of the model.
In our model, we choose the sharp UV cutoff function  $\hat{\vphi}(k)=\chi_{\Lambda}(k)/|k|$, where $\chi_{\Lambda}(k)=1$ if 
$|k|\le \Lambda$, $\chi_{\Lambda}(k)=0$ otherwise. In this case, the
 analysis of the Hamiltonian is  more complicated.
Moreover our focus of interest is not only to prove  the above theorems but also to show  the effectiveness of some new 
operator inequalities in the study  of the nonrelativistic quantum field theory. 

}
\end{rem} 

\section{Monotonically decreasing self-adjoint operators}\label{SecMono}
\setcounter{equation}{0}

In this section, we will provide a strategy of the proof of Theorem
\ref{EnergyMonoWeak} as an abstract theorem.
Through our arguments, it is revealed that essential point of the proof
of Theorem \ref{EnergyMonoWeak} is the operator monotonicity expressed
as (\ref{Monotonicity}).

\subsection{Basic definitions}\label{Def}
Let $\h$ be a complex  Hilbert space  and $\Cone$ be a convex cone in
$\h$.
Then $\Cone $ is called to be {\it self-dual} if 
\begin{align}
\Cone=\{x\in \h\, |\, \la x, y\ra \ge 0\  
\forall y\in \Cone
\}.
\end{align} 
A typical example of self-dual cone is the standard positive cone in
 $L^2(\BbbR^d)$ given by
 $L^2(\BbbR^d)_+=\{f\in
L^2(\BbbR^d)\, |\, f(x)\ge 0 \ \mbox{a.e. $x$}\}$.
Henceforth $\Cone$ always  denotes  the self-dual cone in $\h$.
The following properties of $\Cone$ are well-known \cite{Bos, Haagerup}:
\begin{Prop}\label{BasisSAC}One has the following.
\begin{itemize}
\item[{\rm (i)}] $\Cone\cap (-\Cone)=\{0\}$.
\item[{\rm (ii)}] There exists a unique involution $j$ in $\h$ such that
                 $jx=x$ for all $x\in \Cone$.
\item[{\rm (iii)}] Each element $x\in \h$ with $jx=x$ has a unique
                 decomposition $x=x_+-x_-$,  where $x_+,x_-\in\Cone$ and
                 $\la x_+, x_-\ra=0$.
\item[{\rm (iv)}] $\h$ is linearly spanned by $\Cone$.
\end{itemize} 
\end{Prop} 
If $x-y \in \Cone$, then we will write $x\ge y$ (or $y\le x$)
 w.r.t. $\Cone$.
Let $A$ and $B$ be densely defined linear operators on $\h$.
If $Ax \ge Bx$ w.r.t. $\Cone $ for all $x\in \D(A)\cap \D(B)\cap \Cone$,
then we will write $A\unrhd B$ (or $B\unlhd A$) w.r.t. $\Cone$.
Especially if $A$ satisfies $0\unlhd A$  w.r.t. $\Cone$, then we say
that $A$ {\it preserves positivity with respect to }$\Cone$.
We remark that this symbol ``$\unrhd$'' was first introduced by Miura
 \cite{IMO, Miura}. See also \cite{KiRo2}.

An element $x$ in $\Cone$ is called to be {\it strictly positive } if
$\la x, y \ra > 0$ for all $y\in \Cone\backslash \{0\}$.
We will  write this as $x>0$ w.r.t. $\Cone$. Of course, an inequality  $x>y$
w.r.t. $\Cone $ means $x-y$ is strictly positive w.r.t. $\Cone$.
If bounded operators $A$ and $B$ satsify $Ax > Bx$ w.r.t. $\Cone$ for
all $x\in \Cone\backslash\{0\}$, then we will express this as $A\rhd B$
(or $B\lhd A$) w.r.t. $\Cone$. Clearly if $A\rhd B$ w.r.t. $\Cone$, then
$A\unrhd B$ w.r.t. $\Cone$. We say that $A$ {\it improves positivity
w.r.t. $\Cone$} if $A\rhd 0$ w.r.t. $\Cone$. 

Fundamental properties  of these inequalities  are reviewed  in Appendix \ref{SDCA}.

\subsection{Abstract theorem}

Let $\Cone$ be a self-dual cone in the Hilbert space $\h$.
Let $\{H_n\}_{n\in \BbbN}$  be a sequence of semibounded self-adjoint operators on
$\h$. In this subsection we always assume the following.

\begin{itemize}

\item[{ \bf(A. 1)}] There exists a unique semibounded self-adjoint operator $H$
such that $H_n$ converges to $H$ in strong resolvent sense.
\item[{\bf (A. 2)}] Each $H_n$ has a common domain
	     $\mathscr{D}$. 
\item[{\bf (A. 3)}] For all $n\in \BbbN$ and $s \ge 0$,
$\displaystyle 
\ex^{-s H_n } \unrhd 0
$ w.r.t. $\Cone$. 
\end{itemize}

\begin{Thm}\label{EnergyMono}
Assume (A. 1), (A. 2), (A. 3).
In addition assume
\begin{align}
H_1\unrhd H_2\unrhd\cdots\unrhd H_{n}\unrhd H_{n+1}\unrhd \cdots  \ \ \ \ \ \mbox{w.r.t. $\Cone$.}\label{Monotonicity}
\end{align}
Set $E_n=\inf \mathrm{spec}(H_n)$ and $E=\inf \mathrm{spec}(H)$.
Then  $\{E_n\}_n$ is monotonically decreasing in $n$:
\begin{align}
E_1\ge E_2\ge \cdots\ge E_n\ge E_{n+1}\ge \cdots \ge E
\end{align} 
and $\displaystyle E=\lim_{n\to \infty}E_n$.
\end{Thm}

\subsection{Proof of Theorem  
 \ref{EnergyMono}}

Let $j$ be the involution in Proposition \ref{BasisSAC}.
Since $\ex^{-tH_n}\unrhd 0$ w.r.t. $\Cone$ for all $n\in \BbbN$,
$H_n$ must be $j$-real in the sense that $H_n j=jH_n$.
From this fact, it follows
\begin{align}
E_n=\inf \Big\{
\la\vphi, H_n\vphi\ra\, \Big|\, \vphi\in \mathscr{D}, \ j\vphi=\vphi,\ \|\vphi\|=1
\Big\}.\label{SpecReal}
\end{align} 
Indeed we observe, since $H_n$ is $j$-real,
\begin{align}
\la\vphi, H_n \vphi\ra=\big\la\Re\vphi, H_n\Re \vphi\big\ra+\big\la
 \Im\vphi, H_n\Im
 \vphi\big\ra, \label{ExpReIm}
\end{align} 
where $\Re\vphi=\frac{1}{2}(\one+j)\vphi,\ \Im \vphi=\frac{1}{2\im }
(\one-j)\vphi$. 
Clearly $j\Re\vphi= \Re\vphi$, $j\Im \vphi=\Im\vphi$ and  $\|\vphi\|^2=\|\Re\vphi\|^2+\|\Im\vphi\|^2$.
Hence 
\begin{align}
E_n\ge \mbox{RHS of (\ref{SpecReal})}
\end{align} 
 holds. The converse inequality is trivial.

Fix $\vepsilon>0$ arbitrarily.
By (\ref{SpecReal}), we can choose $\vphi\in \mathscr{D}$ so that
$j\vphi=\vphi$ and $\la \vphi, H_n\vphi\ra\le E_n+\vepsilon$.
Remark that, by Proposition \ref{BasisSAC} (iii), we can express $\vphi$
as $\vphi=\vphi_+-\vphi_-$ so that $\vphi_{\pm}\in \Cone$.
Define $|\vphi|_{\Cone}=\vphi_++\vphi_-$.
Applying Theorem \ref{PPEquiv}, one obtains both $\la |\vphi|_{\Cone}, H_n |\vphi|_{\Cone}\ra
$
and $\la |\vphi|_{\Cone}, H_{n+1} |\vphi|_{\Cone}\ra
$ are finite,
and 
\begin{align}
\la \vphi, H_n\vphi\ra&\ge\la |\vphi|_{\Cone}, H_n |\vphi|_{\Cone}\ra\no
&=\la |\vphi|_{\Cone}, H_{n+1} |\vphi|_{\Cone}\ra+\la |\vphi|_{\Cone}, ( H_n-H_{n+1}) |\vphi|_{\Cone}\ra.
\end{align} 
By the monotonicity (\ref{Monotonicity}), $
\la |\vphi|_{\Cone}, ( H_n-H_{n+1}) |\vphi|_{\Cone}\ra\ge0
$
holds. Now we arrive at 
\begin{align}
 E_n+\vepsilon \ge \la |\vphi|_{\Cone}, H_{n+1}|\vphi|_{\Cone}\ra\ge E_{n+1}.
\end{align}
Since $\vepsilon >0$ is arbitrary, one obtains the assertion. $\Box$

\section{Second quantization}\label{2ndQuant}
\setcounter{equation}{0}
\subsection{Basic definitions}

Here we give some basic definitions of the second quantized operators \cite{BraRob2}.
The bosonic  Fock space over $\h$ is defined by 
\begin{align}
\Fock(\h)=\Sumoplus \h^{\otimes_{\mathrm{s}}n},
\end{align} 
where $\h^{\otimes_{\mathrm{s}}n}$  is the $n$-fold
symmetric tensor product of $\h$ with convention 
$\h^{\otimes_{\mathrm{s}}0}=\BbbC$. The vector
$\Omega=1\oplus 0 \oplus0\oplus  \cdots\in \Fock(\h)$ is called the Fock vacuum.
For each $n\in \{0\}\cup \BbbN$, let $P_n$ be an orthogonal projection  defined by 
$P_n \vphi=\sum_{n\ge j \ge 0}^{\oplus} \vphi_j$ for all
$\vphi=\sum_{j\ge 0}^{\oplus}\vphi_j\in \Fock$. Then
an important dense subspace of $\Fock(\h)$, called the finite particle
subspace,  is defined  by
\begin{align}
\Ffin(\h)=\bigcup_{n\ge 0} P_n \Fock(\h).\label{FiniteSpace}
\end{align}

We denote by $a(f)\, (f\in \h)$ the annihilation operator on
$\Fock(\h)$, its adjoint $a(f)^*$, called the creation operator, is defined by
\begin{align}
a(f)^*\vphi=\sideset{}{^{\oplus}_{n\ge 1}}\sum
 \sqrt{n}S_n (f\otimes \vphi^{(n-1)})\label{DefCrea}
\end{align} 
for $\vphi=\sum_{n\ge 0}^{\oplus} \vphi^{(n)}\in \D(a(f)^*)$, where
$S_n$ is the symmetrizer  on $\h^{\otimes_{\mathrm{s}}n }$.
It is well-known that the annihilation- and creation operators satisfy
the canonical commutation relations or CCRs
\begin{align}
[a(f), a(g)^*]=\la f, g\ra, \ \ [a(f), a(g)]=0=[a(f)^*, a(g)^*]
\end{align} 
on $\Ffin$.

Let $C$ be a contraction operator on $\h$, that is ,
$\|C\|\le 1$. Then we define a contraction operator $\Gamma(C)$ on
$\Fock(\h)$
by 
\begin{align}
\Gamma(C)=\Sumoplus C^{\otimes n}
\end{align} 
with $C^{\otimes 0}=\one$, the identity operator. 
For a self-adjoint operator $A$ on $\h$, let us introduce 
\begin{align}
\dG(A)=0\oplus \sideset{}{^{\oplus}_{n\ge 1}}\sum
 \sideset{}{_{n\ge k\ge 1}}\sum
\one^{\otimes (k-1)}\otimes A \otimes \one^{\otimes (n-k)}  
\end{align} 
acting in   $\Fock(\h)$. Then  $\dG(A)$ is essentially self-adjoint. We denote its
closure by the same symbol. A typical example is the boson number operator $\Nf=\dG(\one)$.
We remark the following relation between
$\Gamma(\cdot)$ and $\dG(\cdot)$:
\begin{align}
\Gamma(\ex^{\im t A})= \ex^{-\im t \dG(A)}.
\end{align} 
In particular if $A$ is positive, then one has 
\begin{align}
\Gamma(\ex^{-tA})=\ex^{-t \dG(A)}.
\end{align} 

Let $A$ be a positive self-adjoint operator. Then the following operator inequalities are well-known:
\begin{align}
a(f)^*a(f)&\le \|A^{-1/2}f\|^2 (\dG(A)+\one),\label{CreAnnInq}\\
\dG(A)+a(f)+a(f)^*&\ge -\|A^{-1/2}f\|^2. \label{VHove}
\end{align} 

\subsection{Fock space over $L^2$-space}
In this paper, the bosonic Fock space over $L^2(\BbbR^3_k)=L^2(\BbbR^3,
\dm k)$ will often appear and we simply denote as 
\begin{align}
\Fock=\Fock(L^2(\BbbR^3_k)).
\end{align} 
Also the corresponding finite particle subspace  $\Ffin(L^2(\BbbR^3_k))$
is denoted by $\Ffin$.
The $n$-boson subspace $L^2(\BbbR^3_k)^{\otimes_{\mathrm{s}}n}$ is naturally
identified with $
L^2_{\mathrm{sym}}(\BbbR^{3n})=\big\{
\vphi\in L^2(\BbbR^{3n}_k)\, |\, \vphi(k_1,\dots,
k_n)=\vphi(k_{\sigma(1)}, \dots, k_{\sigma(n)})\ \mbox{a.e. }\forall
\sigma \in \mathfrak{S}_n 
\big\}
$, where $\mathfrak{S}_n$ is the permutation group on a set $\{1, 2,
\dots , n\}$.
Hence 
\begin{align}
\Fock=\BbbC \oplus \sideset{}{^{\oplus}_{n\ge 1}}\sum L^2_{\mathrm{sym}}(\BbbR^{3n}_k).
\end{align}
The annihilation- and creation operators are symbolically expressed as 
\begin{align}
a(f)=\int_{\BbbR^3}\dm k\, \overline{f(k)}a(k),\ \
 a(f)^*=\int_{\BbbR^3}\dm k\, f(k) a(k)^*.
\end{align}
If $\omega$ is a multipilication operator by the function $\omega(k)$,
then $\dG(\omega)$ is formally written as 
\begin{align}
\dG(\omega)=\int_{\BbbR^3_k}\dm k\, \omega(k)a(k)^* a(k).
\end{align}  

\subsection{The Fr\"ohlich cone  in $\Fock$}
In order to discuss the inequalities introduced in \S \ref{SecMono},  
we have to determine a   self-dual cone in $\Fock$.
Here we will introduce  a  natural self-dual cone in $\Fock$ which is 
 suitable for our analysis in later sections. 

Set 
\begin{align}
\Fock^{(n)}_+
=\big\{
\vphi\in L^2(\BbbR^3_k)^{\otimes_{\mathrm{s}}n}\, |\, \la f_1\otimes
 \cdots \otimes f_n, \vphi\ra \ge 0\ \forall f_1, \dots, \forall f_n \in L^2(\BbbR^3_k)_+
\big\}
\end{align} 
with $\Fock_+^{(0)}=\BbbR_+=\{r\in \BbbR\,
|\, r\ge 0\}$.  Then each $\Fock_+^{(n)}$ is a self-dual cone in $L^2(\BbbR^3_k)^{\otimes_{\mathrm{s}}n}$. 
Under  the natural identification
$L^2(\BbbR^3_k)^{\otimes_{\mathrm{s}}n}=L^2_{\mathrm{sym}}(\BbbR^{3n}_k)$,
one sees
\begin{align}
\Fock_+^{(n)}=\big\{
\vphi \in L^2_{\mathrm{sym}}(\BbbR^{3n})\, |\, \vphi(k_1,\dots, k_n)\ge
 0\ \mbox{a.e.}
\big\}.\label{FockIdentify}
\end{align}
Now we define
\begin{align}
\Fock_+=\Sumoplus \Fock^{(n)}_+.
\end{align} 
Again $\Fock_+$ becomes  a self-dual cone in $\Fock$.
\begin{define}{\rm
$\Fock_+$ is referred to as the {\it Fr\"ohlich cone}.
}
\end{define} 

\begin{rem}
{\rm
The Fr\"ohlich cone was introduced by Fr\"ohlich  \cite{JFroehlich1,
 JFroehlich2} to study the quantum field theory.
}
\end{rem} 

We summarize  properties of operators in $\Fock$ below.
\begin{Prop}
\label{PPFockI}
 Let $C$ be a contraction on $L^2(\BbbR^3_k)$.  Then if
$C\unrhd 0$ w.r.t. $L^2(\BbbR^3_k)_+$,   one has 
$\Gamma(C)\unrhd 0$ w.r.t. $\Fock_+$. Especially one has the following.
\begin{itemize}
\item[{\rm (i)}] For a self-adjoint operator $A$, if $\ex^{\im t A}\unrhd 0$
w.r.t. $L^2(\BbbR^3)_+$, then one has $\Gamma(\ex^{\im t A})\unrhd 0$
w.r.t. $\Fock_+$.
\item[{\rm (ii)}]  For a positive  self-adjoint operator $B$, if $\ex^{- t B}\unrhd 0$
w.r.t. $L^2(\BbbR^3)_+$, then one has $\Gamma(\ex^{-t B})\unrhd 0$
w.r.t. $\Fock_+$.
\end{itemize} 
\end{Prop} 
{\it Proof.} For each $f_1, \dots, f_n\in L^2(\BbbR^3)_+$ and $\vphi \in
\Fock_+$, one can check that 
\begin{align}
\la \Gamma(C)\vphi, f_1\otimes \cdots \otimes f_n\ra=\la \vphi,
 Cf_1\otimes \cdots \otimes Cf_n\ra \ge 0.
\end{align} 
This means $\Gamma(C)\unrhd 0$ w.r.t. $\Fock_+$. $\Box$

\begin{Prop}\label{PPFockII}
If $f \ge 0$ w.r.t. $L^2(\BbbR^3_k)_+$, then $a(f)^*\unrhd 0$ and
 $a(f)\unrhd 0$ w.r.t. $\Fock_+$.
\end{Prop} 
{\it Proof.} By (\ref{DefCrea}), for any $g_1, \dots, g_n \in
L^2(\BbbR^3_k)_+$ and $\vphi\in \Fock_+\cap \D(a(f)^*)$, one has
\begin{align} 
\la a(f)^* \vphi, g_1\otimes \cdots \otimes g_n\ra=\sqrt{n} \la f\otimes
 \vphi^{(n-1)}, S_n g_1\otimes \cdots \otimes g_n\ra \ge0.
\end{align} 
This implies $a(f)^*\unrhd 0$ w.r.t. $\Fock_+$. $\Box$

\begin{Prop}\label{ErgoFock}{\rm (Ergodicity)}
For each $f\in L^2(\BbbR^3_k)$, let $\phi(f)$ be a  linear
 operator
 defined by
\begin{align}
\phi(f)=a(f)+a(f)^*.
\end{align} 
If $f >0$ w.r.t. $L^2(\BbbR^3_k)_+$, that is, $f(k)>0$ a.e. $k$, then
$\phi(f)$ is ergodic in the sense  that, for any $x, y\in
 (\Fock_+\cap \Ffin) \backslash \{0\}$, there exists an $n\in \{0\}\cup\BbbN$ such
 that $\la x, \phi(f)^n y\ra >0$.
\end{Prop} 
{\it Proof.}
First we remark that 
if $f\ge 0$ w.r.t. $L^2(\BbbR^3_k)_+$ ,  $\phi(f) \unrhd 0$
w.r.t. $\Fock_+$ by Proposition  \ref{PPFockII}. Moreover 
$\Ffin \subseteq \D(\phi(f)^n)$ for any $n\in \BbbN$.

Write $x, y\in \Fock_+\backslash \{0\}$ as $x=\sum_{n \ge 0}^{\oplus}
x^{(n)}$ and  $y=\sum_{n \ge 0}^{\oplus}
y^{(n)}$.
Each $x^{(n)}$ and $y^{(n)}$ are in $\Fock_+^{(n)}$.
 Since  both $x$ and $y$ are nonzero vecors in $\Fock_+$, there
exist $p, q \in \{0\}\cup \BbbN$ so that $x^{(p)} \in
\Fock_+^{(p)}$ and  
$y^{(q)} \in \Fock^{(q)}_+$.
Clearly $x \ge \sum_{n\ge 0}^{\oplus} \delta_{np} x^{(n)}$ w.r.t. $
\Fock_+$ and   $y \ge \sum_{n\ge 0}^{\oplus} \delta_{nq} y^{(n)}$ w.r.t. $
\Fock_+$, where $\delta_{mn}$ is Kronecker delta. Hence one has 
\begin{align}
\la x, \phi(f)^{p+q}y\ra \ge \la x^{(p)}, \phi(f)^{p+q} y^{(q)}\ra. \label{LowerPhi}
\end{align} 
Set $\phi_-(f)=a(f)$ and
$\phi_+(f)=a(f)^*$.
Then since $\phi_{\pm}(f)\unrhd 0$ w.r.t. $\Fock_+$ provided that $f\in
L^2(\BbbR^3_k)_+$, we have $\phi(f)^{p+q}\unrhd \phi_+(f)^p \phi_-(f)^{q}$
w.r.t. $\Fock_+$. Accordingly one has 
\begin{align}
\phi(f)^p x^{(p)}& \ge \phi_-(f)^p x^{(p)}=\sqrt{p!}\la
 f^{\otimes p}, x^{(p)}\ra \Omega,\\
\phi(f)^q y^{(q)}& \ge \phi_-(f)^q y^{(q)}=\sqrt{q!}\la
 f^{\otimes q}, y^{(q)}\ra \Omega
\end{align} 
w.r.t. $\Fock_+$. By the assumption $f >0$ w.r.t. $L^2(\BbbR^3_k)_+$, 
$\la f^{\otimes p}, x^{(p)}\ra >0$ and $\la f^{\otimes q}, y^{(q)}\ra
>0$
hold.
Hence  we arrive at, by (\ref{LowerPhi}),
\begin{align}
\la x, \phi(f)^{p+q} y\ra \ge \sqrt{p!q!}\la f^{\otimes p},
 x^{(p)}\ra \la f^{\otimes q}, y^{(q)}\ra>0.\label{Notationphi}
\end{align}
This proves the assertion. $\Box$

\subsection{Local properties}
Let $B_{\Lambda}$ be a ball of radius $\Lambda$ in $\BbbR^3_k$ and let
$\chi_{\Lambda}$ be a function on $\BbbR^3$ defined by
$\chi_{\Lambda}(k)=1$ if $k \in B_{\Lambda}$ and $\chi_{\Lambda}(k)=0$
otherwise. 
Then as a multiplication operator, $\chi_{\Lambda}$
is an orthogonal projection on $L^2(\BbbR^3_k)$ and
 $Q_{\Lambda}=\Gamma(\chi_{\Lambda})$ is also an orthogonal projection  on $\Fock$.
Now let us define the local Fock space by
\begin{align}
\Fock_{\Lambda}=Q_{\Lambda}\Fock.
\end{align} 
Clearly $\Fock=\Fock_{\infty}$. Since $\chi_{\Lambda}L^2(\BbbR^3_k)=L^2(B_{\Lambda})$, $\Fock_{\Lambda}$ can be
identified with $\Fock(L^2(B_{\Lambda}))$. 
As to the annihilation- and creation operators, we remark the following
properties:
\begin{align}
a(f)Q_{\Lambda}&=a(\chi_{\Lambda}f)=\int_{|k|\le \Lambda}\dm k\,
 \overline{f(k)} a(k),\\
Q_{\Lambda}a(f)^*&=a(\chi_{\Lambda}f)^*=\int_{|k|\le \Lambda}\dm k\,
 f(k) a(k)^*,\\
\dG(\omega)Q_{\Lambda}&=\dG(\chi_{\Lambda}\omega)=\int_{|k|\le \Lambda}\dm k\, \omega(k)a(k)^*a(k).
\end{align}

A natural self-dual cone in $\Fock_{\Lambda}$ would be the following
one. First let us define
\begin{align}
\Fock_{\Lambda, +}^{(n)}=\big\{
\vphi\in L^2(B_{\Lambda})^{\otimes_{\mathrm{s}}n}\, |\, \la f_1\otimes
 \cdots \otimes f_n, \vphi\ra \ge 0\ \forall f_1, \dots, \forall f_n\in L^2(B_{\Lambda})_+
\big\}
\end{align}
with $\Fock_{\Lambda, +}^{(0)}=\BbbR^+$, where $L^2(B_{\Lambda})_+=\{f\in
L^2(B_{\Lambda})\, |\, f(k)\ge 0\ \mbox{a.e. }\}$. Then we introduce 
\begin{align}
\Fock_{\Lambda, +}=\Sumoplus \Fock_{\Lambda, +}^{(n)}.
\end{align} 
$\Fock_{\Lambda, +}$ is a self-dual cone in $\Fock_{\Lambda}$.

\begin{define}
{\rm
$\Fock_{\Lambda, +}$ is referred to as the {\it local  Fr\"ohlich cone}.
}
\end{define} 

\begin{Prop}\label{LocalPropErgo}
Propositions \ref{PPFockI}, \ref{PPFockII} and \ref{ErgoFock} are still
 true even if one replaces $L^2(\BbbR^3_k)_+$ and $\Fock_+$ by
 $L^2(B_{\Lambda})_+$ and $\Fock_{\Lambda, +}$ respectively. 
\end{Prop}

\section{Proof of Theorem \ref{EnergyMonoWeak}} \label{ProofEMW}
\setcounter{equation}{0}

\subsection{Reduction}

Our strategy of the proof of Theorem \ref{EnergyMonoWeak} is simple:
we just apply Theorem \ref{EnergyMono}. Thus what we have to do 
is to check every assumptions in Theorem \ref{EnergyMono}.

The assumption (A. 1) is satisfied by Proposition \ref{LimitHami}.
 (A. 2) is satisfied as well,  because $\D(H_{\Lambda}(P))=\D(\Nf)\cap
 \D(\Pf^2)$ for each $\Lambda>0$. This is  an immediate consequence of the Kato-Rellich theorem.
Therefore it suffices to show the following two propositions.

\begin{Prop}\label{PositivityPre}
For all $P\in \BbbR^3, \ \Lambda>0$ and $s\ge 0$,  we have $\ex^{-s
 H_{\Lambda}(P)}\unrhd 0$
 w.r.t. $\Fock_+$.
\end{Prop}
The above  proposition  corresponds to the assumption (A. 3).
Next propositon means the assumption (\ref{Monotonicity}) is fulfilled.
\begin{Prop} \label{PolaronUVMono}
For each $P\in \BbbR^3$,
$\{H_{\Lambda}(P)\}_{\Lambda}$ is monotonically decreasing in a sense 
that if $\Lambda\le\Lambda'$, then $H_{\Lambda}(P)\unrhd H_{\Lambda'}(P)$
w.r.t. $\Fock_+$.
\end{Prop} 

In the remainder of this section, we will show two propositions above.

\subsection{Proof of Proposition  \ref{PositivityPre}}

Let us write the Hamiltonian $H_{\Lambda}(P)$ as 
\begin{align}
H_{\Lambda}(P)=H_0(P)-V_{\Lambda},
\end{align} 
where 
\begin{align}
H_0(P)=\frac{1}{2}(P-\Pf)^2+\Nf,\ \ \ \ 
V_{\Lambda}=\sqrt{\alpha}\lambda_0\int_{|k|\le \Lambda}
\dm k\, \frac{1}{|k|}[a(k)+a(k)^*].
\end{align} 
 $V_{\Lambda}$ is the electron-phonon interaction term.

\begin{lemm}\label{Attraction}
For all $P\in \BbbR^3$ and $\Lambda>0$, one obtains the following.
\begin{itemize}
\item[{\rm (i)}] $\ex^{-t H_0(P)}\unrhd 0$ w.r.t. $\Fock_+$ for all
	     $t\ge 0$.
\item[{\rm (ii)}]  $-V_{\Lambda}$ is attractive
	     w.r.t. $\Fock_+$
 in a sense $-V_{\Lambda}\unlhd 0$ w.r.t. $\Fock_+$.
\end{itemize} 
\end{lemm} 
{\it Proof.}
(i) By Proposition \ref{PPFockI}, $\ex^{-t \Nf}\unrhd 0$ w.r.t. $\Fock_+$.
Furthermore $\ex^{-t (P-\Pf)^2}\unrhd 0$ w.r.t. $\Fock$ for all $P$.
[Proof:
We can write $\ex^{-t (P-\Pf)^2}=\ex^{-t |P|^2} \oplus \sum_{n\ge 1}^{\oplus}
\exp\{-t (P-\sum_{j=1}^n k_j)^2\}$. Each $n$-th component satisfies $\exp\{-t (P-\sum_{j=1}^n k_j)^2\} \unrhd 0$
w.r.t. $\Fock_+^{(n)}$.]
 This implies $\exp[-t H_0(P)]
=\exp[-t \frac{1}{2}(P-\Pf)^2]\exp[-t \Nf] \unrhd 0$ w.r.t. $\Fock_+$
for all $P$.

(ii) This immediately follows from Proposition \ref{PPFockII}. $\Box$
\medskip\\

Now we are ready to prove Proposition \ref{PositivityPre}. 
 By Lemma
\ref{Attraction}, every
assumptions in Proposition \ref{PerturbationPP} are proven already.
Thus Proposition \ref{PositivityPre} follows from Proposition
\ref{PerturbationPP}. $\Box$

\subsection{Proof of Propositon \ref{PolaronUVMono}}

First of all, we will clarify a property of $V_{\Lambda}$.

\begin{lemm}\label{DecreasingV}
$-V_{\Lambda}$ is monotonically decreasing in $\Lambda$ in a sense 
$-V_{\Lambda}\unrhd -V_{\Lambda'}$ w.r.t. $\Fock_+$ provided
 $\Lambda\le\Lambda'$.
\end{lemm} 
\begin{rem}
{\rm
By Lemma \ref{DecreasingV}, the electron-phonon interaction becomes 
 stronger, the larger we take the ultraviolet cutoff. 
(Here our choice of the Fr\"ohlich  cone is essential.)
}
\end{rem} 
{\it Proof.} 
Remark that, since $\D(V_{\Lambda})\cap \D(V_{\Lambda'})\supseteq
\Ffin$, we see  $
\D(V_{\Lambda})\cap \D(V_{\Lambda'})\cap \Fock_+ \neq \{0\}.
$
Choose $\vphi\in \D(V_{\Lambda})\cap \D(V_{\Lambda'})\cap \Fock_+$. 
Then, applying   Propositon \ref{PPFockII}, we have 
\begin{align}
(V_{\Lambda'}-V_{\Lambda})\vphi=
\sqrt{\alpha} \lambda_0
\int_{\Lambda<|k|\le \Lambda'}\dm k\,
 \frac{1}{|k|}[a(k)+a(k)^*]\vphi \ge 0 \label{DiffV}
\end{align} 
w.r.t. $\Fock_+$.
This means $V_{\Lambda'}\unrhd V_{\Lambda}$ w.r.t.  $\Fock_+$. $\Box$
\medskip\\

Assume $\Lambda'\ge\Lambda$. For $\vphi\in \D(\Nf)\cap \D(\Pf^2)\cap \Fock_+$,
observe, by Lemma \ref{DecreasingV}, 
\begin{align}
\big(H_{\Lambda}(P)-H_{\Lambda'}(P)\big)\vphi=(V_{\Lambda'}-V_{\Lambda})\vphi \ge 0
\end{align} 
w.r.t. $\Fock_+$.
Since $\D(\Nf)\cap \D(\Pf^2)$ is the  common domain of $H_{\Lambda}(P)$
and $H_{\Lambda'}(P)$, we conclude  $H_{\Lambda}(P)\unrhd H_{\Lambda'}(P)$ w.r.t. $\Fock_+$ for all
$P\in \BbbR^3$. This proves Proposition \ref{PolaronUVMono}.
$\Box$

\section{Proof of Theorem \ref{EnergyMonoPolaron}} \label{ProofLocalPr}

\subsection{Local Hamiltonian}
By the factorization $\Fock(\h_0\oplus
\h_1)=\Fock(\h_0)\otimes \Fock(\h_1)$, one has 
\begin{align}
\Fock=&\Fock(L^2(B_{\Lambda})\oplus L^2(B_{\Lambda}^{\mathrm{c}}))
= \Fock(L^2(B_{\Lambda}))\otimes
 \Fock(L^2(B_{\Lambda}^{\mathrm{c}}))\no
=&\Sumoplus \Fock_{\Lambda}\otimes
 L^2_{\mathrm{sym}}(B_{\Lambda}^{\mathrm{c}\times n})
=\Fock_{\Lambda}\oplus \sideset{}{^{\oplus}_{n\ge 1}}\sum
 L^2_{\mathrm{sym}}(B_{\Lambda}^{\mathrm{c}\times n}; \Fock_{\Lambda}),\label{FockIdentification}
\end{align} 
where  $L^2_{\mathrm{sym}}(B_{\Lambda}^{\mathrm{c}\times n};  \Fock_{\Lambda})$ is the
space of symmetric square integrable $\Fock_{\Lambda}$-valued functions on
$B_{\Lambda}^{\mathrm{c}\times n}$ and $B_{\Lambda}^{\mathrm{c}}=\BbbR^3\backslash B_{\Lambda}$. Under this identification, we see
that 
\begin{align}
H_{\Lambda}(P)=K_{\Lambda}(P)\oplus  \sideset{}{^{\oplus}_{n\ge 1}}\sum
\int^{\oplus}_{B_{\Lambda}^{\mathrm{c}\times n}} \Big[
K_{\Lambda}(P-k_1-\cdots-k_n)+n\Big]\, \dm k_1\cdots \dm
 k_n.\label{HDecomposition}
\end{align} 
Here $K_{\Lambda}(P)$ is the {\it local Hamiltonian} defined by 
\begin{align}
K_{\Lambda}(P)=\frac{1}{2}(P-P_{\mathrm{f}, \Lambda})^2+N_{\mathrm{f},
 \Lambda}
-V_{\Lambda}, \label{reduction}
\end{align} 
where 
\begin{align}
P_{\mathrm{f}, \Lambda}=\int_{|k|\le \Lambda}\dm k\, k a(k)^* a(k),\ \ \ 
N_{\mathrm{f}, \Lambda}=\int_{|k|\le \Lambda}\dm k\, a(k)^*a(k).
\end{align} 
$K_{\Lambda}(P)$ lives in the local Fock space $\Fock_{\Lambda}$. By the
Kato-Rellich theorem, it is self-adjoint on $\D(P_{\mathrm{f},
\Lambda}^2)\cap\D(N_{\mathrm{f}, \Lambda})$. 

Put 
\begin{align*}
L(P)=\frac{1}{2}(P-P_{\mathrm{f}, \Lambda})^2+N_{\mathrm{f},
 \Lambda}.
\end{align*} 
Obviously $K_{\Lambda}(P)=L(P)-V_{\Lambda}$.

\begin{lemm}\label{LocalAttraction}
For each $\Lambda>0$ and $P\in \BbbR^3$, one has the following.
\begin{itemize}
\item[{\rm (i)}] $\ex^{-t L(P)}\unrhd 0$ w.r.t. $\Fock_{\Lambda, +}$ for
	     all $t\ge 0$.
\item[{\rm (ii)}] $-V_{\Lambda}$ is attractive
	     w.r.t. $\Fock_{\Lambda, +}$ in a sense that
	     $-V_{\Lambda}\unlhd 0$ w.r.t. $\Fock_{\Lambda, +}.$
\end{itemize} 
\end{lemm} 
{\it Proof.} This can be proven in a similar way in the proof of Lemma
\ref{Attraction}. $\Box$

\begin{coro}
For each $\Lambda>0$ and $P\in \BbbR^3$, one obtains 
$
\ex^{-t K_{\Lambda}(P)}\unrhd 0
$
w.r.t. $\Fock_{\Lambda, +}$ for all $t\ge 0$.
\end{coro} 
{\it Proof.} Apply Proposition  \ref{PerturbationPP}. $\Box$
\medskip\\

As to $\ex^{-t K_{\Lambda}(P)}$,  we can show a stronger result as follow.

\begin{Prop}\label{LocalErgodicity} 
For any $\Lambda>0, P\in \BbbR^3$ and $t>0$, one obtains 
$
\ex^{-t K_{\Lambda}(P)} \rhd 0
$
w.r.t. $\Fock_{\Lambda, +}$.
\end{Prop} 
{\it Proof.} Essential idea comes from \cite{JFroehlich1, JFroehlich2}.
Set $F(k)=\sqrt{\alpha}\lambda_0 \chi_{\Lambda}(k)/|k|$.  Since $F >0$
w.r.t. $L^2(B_{\Lambda})_+$, $V_{\Lambda}$ is ergodic by Proposition
\ref{LocalPropErgo}.

By the Duhamel formula, one observes 
\begin{align}
\ex^{-t K_{\Lambda}(P)}=\sum_{j\ge 0}D_j(t)
\end{align} 
with
\begin{align}
D_j(t)=&\int_0^{t}\dm s_1 \int_0^{t-s_1}\dm s_2 \cdots
 \int_0^{t-\sum_{i=1}^{j-1}s_i}\dm s_j\no &\ex^{-s_1 L(P)}
 V_{\Lambda}\ex^{-s_2 L(P)}\cdots \ex^{- s_j L(P)}V_{\Lambda}\ex^{-(t-\sum_{i=1}^{j}s_i)L(P)}.
\end{align} 
Since each $D_j(t)\unrhd 0$ w.r.t. $\Fock_{\Lambda, +}$ by Lemma \ref{LocalAttraction}, one has $\ex^{-t
K_{\Lambda}(P)}\unrhd D_j(t)
$ w.r.t. $\Fock_{\Lambda, +}$ for any $j$.  Hence it sufficies to show that a sequence
$\{D_j(t)\}_j$ is ergodic in the sense that, for any $\vphi, \psi\in
\Fock_{\Lambda, +}\backslash \{0\}$, there exists some $N\in \{0\}\cup \BbbN $ such
that $\la \vphi, D_N(t)\psi \ra>0$.
To this end, write $\vphi=\sum_{n\ge 0}^{\oplus} \vphi^{(n)}$ and
$\psi=\sum_{n\ge 0}^{\oplus}\psi^{(n)}$. Then since both $\vphi$ and
$\psi$ are non-zero, there are $p, q\in \{0\}\cup \BbbN$ such that 
$\vphi^{(p)}\in \Fock_{\Lambda, +}^{(p)}\backslash\{0\}$ and $\psi^{(q)}\in
\Fock_{\Lambda, +}^{(q)}\backslash \{0\}$.
Since $\vphi \ge \vphi^{(p)}$ and $\psi\ge \psi^{(q)}$
w.r.t. $\Fock_{\Lambda, +}$,
one sees 
\begin{align}
\la \vphi, D_j(t)\psi\ra \ge \la \vphi^{(p)}, D_j(t)\psi^{(q)}\ra\label{DLower}
\end{align} 
for any $j$. Observe that, by the local ergodicity of $V_{\Lambda}$ (Proposition
\ref{LocalPropErgo}), there
exists an $N\in \{0\}\cup \BbbN$, such that 
$\la \vphi^{(p)}, V_{\Lambda}^N \ex^{-t L(P)}\psi^{(q)}\ra >0$. This
implies $\la \vphi^{(p)}, D_N(t)\psi^{(q)}\ra >0$. Hence combining  this
with (\ref{DLower}), $\{D_j(t)\}_j$ is ergodic. $\Box$

\subsection{Proof by the local  operator properties}
We will prove Theorem \ref{EnergyMonoPolaron} by clarifying relations
between $K_{\Lambda}(P)$ and $H_{\Lambda}(P)$.

\begin{lemm}\label{LocalSpectrum}
Let $K_{\Lambda}(P)$ be the local Hamiltonian defined by
 (\ref{reduction}). Let $\mathcal{E}_{\Lambda}(P)=\inf \mathrm{spec}(K_{\Lambda}(P))$.
Then, for $|P|<\sqrt{2}$, one has 
$
\mathcal{E}_{\Lambda}(P)=E_{\Lambda}(P).
$
\end{lemm} 
{\it Proof.} 
Using the  property $\mathcal{E}_{\Lambda}(0)\le
\mathcal{E}_{\Lambda}(P)$ (Lemma \ref{LocalEInq}),
one has 
\begin{align}
E_{\Lambda}(P)=\min\{\mathcal{E}_{\Lambda}(P),
 \mathcal{E}_{\Lambda}(0)+1\} \label{EqHK}
\end{align} 
by (\ref{HDecomposition}). 
On the other hand, we see that 
\begin{align}
\mathcal{E}_{\Lambda}(P)\le \mathcal{E}_{\Lambda}(0)+\frac{P^2}{2}. \label{EnergyInq}
\end{align} 
[Proof:
For each normalized $\vphi$, $\la \vphi,
[K_{\Lambda}(P)-\frac{P^2}{2}]\vphi\ra$
is linear in $P$. Hence $F(P)=\mathcal{E}_{\Lambda}(P)-\frac{P^2}{2}$ is concave.
Now we have $F(0)=F(\frac{P}{2}-\frac{P}{2})\ge
\frac{1}{2}F(P)+\frac{1}{2}F(-P)$. Finally using the fact $F(-P)=F(P)$
which can be proven by, for example, the time reversal symmetry \cite{LMS}, we
conclude (\ref{EnergyInq}).
] Combining (\ref{EqHK}) and (\ref{EnergyInq}), we have the assertion. $\Box$
\medskip\\

By the above lemma, it sufficies to consider the local Hamiltonian
$K_{\Lambda}(P)$ instead of $H_{\Lambda}(P)$.

\begin{lemm}\label{LocalGS}
For any $\Lambda>0$, $K_{\Lambda}(P)$ has a ground state provided   $|P|<\sqrt{2}$.
\end{lemm} 
{\it Proof.} First we recall the following fact:  $H_{\Lambda}(P)$ has a
normalized  ground
state  $\Psi_{\Lambda}(P)$  for $|P|<\sqrt{2}$. As to the proof, see
\cite{JFroehlich1,
GeLowen, Spohn2}.
Corresponding to the decomposition (\ref{FockIdentification}), one can
write 
\begin{align}
\Psi_{\Lambda}(P)=\Psi_{\Lambda}^{(0)}(P)\oplus
 \sideset{}{^{\oplus}_{n\ge 1}}\sum
\Psi_{\Lambda}^{(n)}(P),
\end{align} 
where $\Psi^{(0)}_{\Lambda}(P)\in \Fock_{\Lambda}$ and
$\Psi^{(n)}_{\Lambda}(P)\in
L^2_{\mathrm{sym}}(B_{\Lambda}^{\mathrm{c}\times n}; \Fock_{\Lambda}).
$

Put 
$
\mathcal{S}_{\Lambda, P}=\{
n\in \BbbN\, |\, \Psi_{\Lambda}^{(n)}(P)\neq 0
\}.
$
Assume $\mathcal{S}_{\Lambda, P}$ is not empty.
 The we have $1=\|\Psi_{\Lambda}(P)\|^2=\|\Psi_{\Lambda}^{(0)}(P)\|^2+
\sum_{n\in \mathcal{S}_{\Lambda, P}}\|\Psi_{\Lambda}^{(n)}(P)\|^2
$
and, by (\ref{HDecomposition}) and Lemma \ref{LocalEInq},
\begin{align}
E_{\Lambda}(P)&=\la \Psi_{\Lambda}(P),
 H_{\Lambda}(P)\Psi_{\Lambda}(P)\ra\no
&\ge \mathcal{E}_{\Lambda}(P)\|\Psi_{\Lambda}^{(0)}(P)\|^2+
\sum_{n \ge 1} \big[
\mathcal{E}_{\Lambda}(0)+1
\big]\|\Psi_{\Lambda}^{(n)}(P)\|^2\no
&= \mathcal{E}_{\Lambda}(P)\|\Psi_{\Lambda}^{(0)}(P)\|^2+
\sum_{n\in \mathcal{S}_{\Lambda, P}}
\big[
\mathcal{E}_{\Lambda}(0)+1
\big]\|\Psi_{\Lambda}^{(n)}(P)\|^2. \label{EVepsilon}
\end{align} 
By (\ref{EnergyInq}), one has 
$
\big[
\mathcal{E}_{\Lambda}(0)+1
\big]-\mathcal{E}_{\Lambda}(P)
\ge 1-\frac{P^2}{2}>0
$
provided $|P|<\sqrt{2}$. Thus, if $|P|<\sqrt{2}$, 
$
\mbox{RHS of (\ref{EVepsilon})}>\mathcal{E}_{\Lambda}(P)
$ holds.
This contradicts with (\ref{EqHK}). Hence $\mathcal{S}_{\Lambda,
P}=\emptyset$
and $\Psi_{\Lambda}(P)=\Psi_{\Lambda}^{(0)}(P)\oplus
0\oplus 0\oplus\cdots $. Moreover $\Psi_{\Lambda}^{(0)}(P)$ must be a
ground state of $K_{\Lambda}(P)$. $\Box$
\medskip
\\

\begin{coro}For each $|P|<\sqrt{2}$ and $\Lambda>0$,
the ground state
$\Psi_{\Lambda}^{(0)}(P)$  of $K_{\Lambda}(P)$ is unique in $\Fock_{\Lambda}$
and can be chosen strictly positive w.r.t.  $\Fock_{\Lambda, +}$.
\end{coro} \label{LocalUnique}
{\it Proof.}
This immediately follows from  the local ergodicity (Proposition
\ref{LocalErgodicity}) and Theorem \ref{Faris}. $\Box$
\medskip\\

Next we regard $K_{\Lambda}(P)$ as a self-adjoint operator on a larger  subspace
 $\Fock_{\Lambda'}$ for $\Lambda'>\Lambda$. Then $\Psi_{\Lambda}^{(0)}(P)$ can be regarded as a vector in
$\Fock_{\Lambda', +}$. 
\begin{lemm}\label{NotGS}
Let $\Psi_{\Lambda}^{(0)}(P)$ be the unique ground state of $K_{\Lambda}(P)$.
Then, for each $|P|<\sqrt{2}$ and $\Lambda'>\Lambda$,
 $\Psi_{\Lambda}^{(0)}(P)$ is {\bf not} the ground state of $K_{\Lambda'}(P)$.
\end{lemm} 
{\it Proof.}
Note that $\Psi_{\Lambda}^{(0)}(P)$ is not strictly positive
w.r.t. $\Fock_{\Lambda', +}$ anymore (but it is still positive
w.r.t. $\Fock_{\Lambda', +}$). On the other hand,  the ground state of
$K_{\Lambda'}(P)$ must be unique and strictly positive by Proposition
\ref{LocalErgodicity} and Theorem \ref{Faris}. However since 
$\Psi_{\Lambda}^{(0)}(P)$ is not strictly
positive,  it can not be the ground state of $K_{\Lambda'}(P)$.
$\Box$ 
\medskip\\

If $\Lambda'>\Lambda$, then one has 
$Q_{\Lambda}K_{\Lambda'}(P)Q_{\Lambda}=Q_{\Lambda}K_{\Lambda}(P)Q_{\Lambda}
$. Since
 $\Psi_{\Lambda}^{(0)}(P)$ is not
the ground state of  $K_{\Lambda'}(P)$ by Lemma \ref{NotGS},  one has 
\begin{align}
\mathcal{E}_{\Lambda}(P)&=\la \Psi_{\Lambda}^{(0)}(P),
 Q_{\Lambda}K_{\Lambda}(P)Q_{\Lambda}\Psi_{\Lambda}^{(0)}(P)\ra\no
&=\la
 \Psi_{\Lambda}^{(0)}(P),
 Q_{\Lambda}K_{\Lambda'}(P)Q_{\Lambda}\Psi_{\Lambda}^{(0)}(P)\ra\no
 &>
 \mathcal{E}_{\Lambda'}(P).\label{LocalEngInq}
\end{align} 
Indeed, suppose  
$
\la
 \Psi_{\Lambda}^{(0)}(P),
 Q_{\Lambda}K_{\Lambda'}(P)Q_{\Lambda}\Psi_{\Lambda}^{(0)}(P)\ra =
 \mathcal{E}_{\Lambda'}(P).
$ 
Then $\Psi_{\Lambda}^{(0)}(P)=Q_{\Lambda} \Psi_{\Lambda}^{(0)}(P)$ must be the
ground state of $K_{\Lambda'}(P)$. But this contradicts Lemma
\ref{NotGS} so that the last  inequality in (\ref{LocalEngInq}) holds
.
Combining (\ref{LocalEngInq}) with Lemma \ref{LocalSpectrum}, one arrives at  the desired assertion
in Theorem \ref{EnergyMonoPolaron}. $\Box$

\appendix

\section{Fundamental properties of the operator inequalities }\label{SDCA}
\setcounter{equation}{0}
In this section, we will review some preliminary results about the
inequalities introduced in \S \ref{Def}. Almost all of results here are
taken from the author's previous work \cite{Miyao, Miyao2}.

\subsection{Basic tools }
Let $\mathfrak{v}$ be a dense subspace of the Hilbert space $\h$.
Set 
\begin{align}
\mathcal{L}(\mathfrak{v})=\Big\{
A: \mbox{linear operator on $\h$ s.t. $\mathfrak{v}\subseteq \D(A),
 A\mathfrak{v}\subseteq \mathfrak{v}, A^* \mathfrak{v}\subseteq \mathfrak{v}
$ }
\Big\}.
\end{align} 
Obviously $\mathcal{L}(\mathfrak{v})$ is a linear space and closed under
the operator product, that is, if $A, B\in \mathcal{L}(\mathfrak{v})$,
then $AB\in \mathcal{L}(\mathfrak{v})$.
In this subsection, we always assume every operator in the lemmas
belongs to $\mathcal{L}(\mathfrak{v})$. This tacit assumption remove
unnecessary complexities on domain problem. For instance, the abnormal
case $\Cone\cap \D(A)=\{0\}$ can be avoided automatically. We remark that all
operators in the main sections actually satisfy the assumption under a
suitable choice of $\mathfrak{v}$.

The following two lemmata are immedate consequences of the definitions.

\begin{lemm}\label{PPInq}Suppose that  $0\unlhd A_1\unlhd B_1$ and $0\unlhd A_2\unlhd
 B_2$ w.r.t $\Cone$.
Then one has the following.
\begin{itemize}
\item[{\rm (i)}]   $0\unlhd A_1A_2$ w.r.t. $\Cone$. Moreover if $A_1,
		   B_1\in\mathfrak{B}(\h)$,  the set of all bounded operators on
 $\h$, 
 then $0\unlhd A_1A_2\unlhd B_1 B_2$ w.r.t. $\Cone$.
\item[{\rm (ii)}]  $0\unlhd a A_1+b A_2 \unlhd a B_1+bB_2$ w.r.t. $\Cone$, for all
                   $a,b\in\BbbR_+=\{x\in\BbbR\, |\, x\ge 0\}$.
\item[{\rm (iii)}] Let $A$ be positivity preserving: $0\unlhd
                   A$ w.r.t. $\Cone$. Suppose that $\Cone\cap \D(A)$ is dense in
                   $\Cone$. Then 
                    $0\unlhd A^*$ w.r.t. $\Cone$.
\end{itemize}
\end{lemm}

\begin{lemm}\label{PIBasic}
Let $A, B\in\mathfrak{B}(\h)$. Suppose that 
 $0\lhd A$ and $0\unlhd B$
 w.r.t. $\Cone$. Then  we have the following properties.
\begin{itemize}
\item[{\rm (i)}] $0\lhd A^*$ w.r.t. $\Cone$.
\item[{\rm (ii)}] Suppose that $\ker B^{\#}=\{0\}$ with $a^{\#}=a$ or
		 $a^*$.  Then $0\lhd AB$ and $0\lhd BA$ w.r.t. $\Cone$.
\item[{\rm (iii)}] $0\lhd aA+bB$ w.r.t. $\Cone$ for $a>0$ and $b\ge 0$.
\end{itemize}
\end{lemm}

\subsection{Operator monotonicity}

\begin{Prop}{\rm (Monotonicity)}\label{ABIneqEq}
Let $A$ and $B$ be positive self-adjoint operators. We assume the following.
\begin{itemize}
\item[{\rm (a)}] $\D(A)\subseteq\D(B)$ or $\D(A)\supseteq \D(B)$.
\item[{\rm (b)}] $(A+s)^{-1}\unrhd 0$  and  $(B+s)^{-1}\unrhd 0$ w.r.t. $\Cone$ for all $s>0$.
\end{itemize} 
Then the following are equivalent to each other.
\begin{itemize}
 \item[{\rm (i)}] $B\unrhd A$ w.r.t. $\Cone$.
 \item[{\rm (ii)}] $(A+s)^{-1}\unrhd (B+s)^{-1}$ w.r.t. $\Cone$ for all $s>0$.
 \item[{\rm (iii)}] $\ex^{-tA}\unrhd \ex^{-tB}$ w.r.t. $\Cone$ for all $t\ge 0$.
 \end{itemize}
\end{Prop}
{\it Proof.} (i) $\Rightarrow$ (ii): By the assumptions (a) and  (b), we see that 
\begin{align*}
(A+s)^{-1}-(B+s)^{-1}=(A+s)^{-1} (B-A) (B+s)^{-1} \unrhd 0.
\end{align*}

(ii) $\Rightarrow$ (iii): 
\begin{align*}
\ex^{-tA}=\mbox{s-}\lim_{n\to \infty} (\one+tA/n)^{-n}
\unrhd\mbox{s-}\lim_{n\to \infty} (\one+tB/n)^{-n}= \ex^{-tB}.
\end{align*}

(iii) $\Rightarrow$ (i): 
\begin{align*}
A=\mbox{s-}\lim_{t\downarrow 0} (\one-\ex^{-tA})/t \unlhd \mbox{s-}\lim_{t\downarrow 0} (\one-\ex^{-tB})/t=B. \ \ \Box
\end{align*}

\begin{Prop}\label{PerturbationPP}
Let $A$ be a positive self-adjoint operator and let $B$ be a  symmetric
 operator. Assume the following.
\begin{itemize}
\item[{\rm (i)}] $B$ is $A$-bounded with relative bound $a<1$, i.e.,
                 $\D(A)\subseteq \D(B)$ and $\|Bx\|\le a \|Ax\|+b\|x\|$
                 for all $x\in \D(A)$.
\item[{\rm (ii)}] $0\unlhd \ex^{-tA}$ w.r.t. $\Cone$ for all $t\ge 0$.
\item[{\rm (iii)}]$0\unlhd -B$ w.r.t. $\Cone$.
\end{itemize} 
Then $ \ex^{-t(A+B)}\unrhd \ex^{-tA}\unrhd 0$ w.r.t. $\Cone$ for all $t\ge 0$.
\end{Prop}
{\it Proof.}  See \cite{Miyao}.
$\Box$

\subsection{Beurling-Deny criterion}

Let $j$ be the involution  given in \S \ref{BasisSAC}.
Let $A$ be a linear operator acting in $\h$.
We say that $A$ is {\it $j$-real} if $j\D(A)\subseteq \D(A)$ and $jAx=Ajx$ for
all $x\in \D(A)$. Set $\h_{\mathbb{R}}=\{x\in \h\,
|\, jx=x\}$. Then for any $x\in \h_{\mathbb{R}}$, we have a
unique decomposition $x=x_+-x_-$ with $x_{\pm}\in \Cone$ and $\la x_+,
x_-\ra=0$. Recall the notation $|x|_{\Cone}=x_++x_-$.

The following theorem is an abstract version of Beurling-Deny
criterion \cite{BD}. 

\begin{Thm}\label{PPEquiv}{\rm (Beurling-Deny criterion)}
Let $A$ be a  positive self-adjoint operator on $\h$.
Assume that $A$ is $j$-real.
 Then the
 following are equivalent.
\begin{itemize}
\item[{\rm (i)}]$0\unlhd \ex^{-tA}$ for all $t\ge 0$.
\item[{\rm (ii)}] If $x\in\D(A)\cap \h_{\mathbb{R}}$, then $|x|_{\Cone}\in
                \D(A^{1/2})\cap\h_{\mathbb{R}}$ and $
 \la |x|_{\Cone}, A|x|_{\Cone}\ra\le \la x, A x\ra.$
\item[{\rm (iii)}] If $x\in\D(A)\cap \h_{\mathbb{R}}$, then $x_+\in
                \D(A^{1/2})\cap\h_{\mathbb{R}}$ and $\la x_+, Ax_+\ra\le \la x, A
                x\ra$.
\item[{\rm (iv)}]  If $x\in\D(A)\cap \h_{\mathbb{R}}$, then $x_{\pm}\in
                \D(A^{1/2})\cap\h_{\mathbb{R}}$ and $\la x_+, Ax_+\ra+\la x_-, Ax_-\ra\le \la x, A
                x\ra$.
\end{itemize}
\end{Thm}
{\it Proof.}  Proof is a slight modification of \cite[Theorem XIII.50]{ReSi4}.  $\Box$

\subsection{Perron-Frobenius-Faris theorem}

\begin{Thm}\label{Faris}{\rm (Perron-Frobenius-Faris)}
Let $A$ be a positive self-adjoint operator on $\h$. Suppose that 
 $0\unlhd \ex^{-tA}$ w.r.t. $\Cone$ for all $t\ge 0$ and $\inf
 \mathrm{spec}(A)$ is an eigenvalue.
Let $P_A$ be the orthogonal projection onto the closed subspace spanned
 by  eigenvectors associated with   $\inf
 \mathrm{spec}(A)$.
 Then the following
 are equivalent.
\begin{itemize}
\item[{\rm (i)}] 
$\dim \ran P_A=1$ and $P_A\rhd 0$ w.r.t. $\Cone$.
\item[{\rm (ii)}] $0\lhd (A+s)^{-1}$ for some $s>0$ w.r.t. $\Cone$.
\item[{\rm (iii)}] For all $x,y\in\Cone\backslash \{0\}$, there exists a 
  $t> 0$ such that 
$0<\la x,\ex^{-tA}y\ra$.
\item[{\rm (iv)}] $0\lhd (A+s)^{-1}$ for all $s>0$ w.r.t. $\Cone$.
\item[{\rm (v)}] $0\lhd \ex^{-tA}$ for all $t>0$ w.r.t. $\Cone$.
\end{itemize}
\end{Thm} 
{\it Proof.} See, e.g., \cite{Faris, Miyao, ReSi4}. $\Box$

\section{An energy inequality}\label{LocalEnergyInq}

\begin{lemm}\label{LocalEInq}
For all $P\in \BbbR^3$ and $0<\Lambda<\infty$, one has 
\begin{align}
\mathcal{E}_{\Lambda}(0)\le \mathcal{E}_{\Lambda}(P).\label{LocalEnergy}
\end{align} 
\end{lemm} 
{\it Sketch of proof.}
Since we need  a special self-dual cone different from $\Fock_+$,
we separate  the proof of (\ref{LocalEnergy}) from the main body.

In this appendix, we switch our representation space to the $Q$-space 
or the Schr\"odinger representation. In this representation,
the local Fock space $\Fock_{\Lambda}$ can be identified with
$L^2(Q_{\Lambda})=L^2(Q_{\Lambda}, \dm \mu_{\Lambda})$,
 where $\mu_{\Lambda}$ is a Gaussian measure, see \cite{Simon} for
 details. Let
\begin{align}
L^2(Q_{\Lambda})_+=\{F\in L^2(Q_{\Lambda})\, |\, F\ge 0\, \mbox{a.e.}\}.
\end{align} 
Clearly $L^2(Q_{\Lambda})_+$ is a self-dual cone in $L^2(Q_{\Lambda})$.
The conjugation $C$ in the one particle space is given by
$(Cf)(k)=\overline{f}(-k)$.  Then, by a general theorem \cite[Theorem
I. 12 and its remark]{Simon},
one sees
\begin{align}
\ex^{\im a\cdot P_{\mathrm{f}, \Lambda}}\unrhd 0,\ \ \ \ex^{-t
 N_{\mathrm{f}, \Lambda}}\unrhd 0,\ \ \ \ex^{t V_{\Lambda}}\unrhd 0
\end{align} 
w.r.t. $L^2(Q_{\Lambda})_+$, as operators in the $Q$-space.
Hence, following Gross \cite{LGross1}, one has 
\begin{align}
\big|\ex^{-t (P-P_{\mathrm{f}, \Lambda})^2} F\big|
\le \ex^{-t P_{\mathrm{f}, \Lambda}^2}|F|\ \ \mbox{ a.e.}
\end{align} 
for each $F\in L^2(Q_{\Lambda})$.
This implies $|\ex^{-t L(P)}F|\le \ex^{-t L(0)}|F|$ a.e..
By the Trotter-Kato formula, one obtains 
$|\ex^{-t K_{\Lambda}(P)}F| \le \ex^{-t K_{\Lambda}(0)}|F|$ a.e..
From this, it follows
$\la F, \ex^{-t K_{\Lambda}(P)} F\ra\le \la |F|, \ex^{-t
K_{\Lambda}(0)}|F|\ra$ for all $F\in L^2(Q_{\Lambda})$.
Now we arrive at the desired result (\ref{LocalEnergy}). $\Box$

\end{document}